\documentclass[11pt]{article}
\usepackage[latin1]{inputenc}
\usepackage[english]{babel}
\usepackage[namelimits]{amsmath}
\usepackage{amssymb}
\usepackage{amsmath}
\usepackage{amsthm}

\begin{document}
\title{A possible new cosmological redshift effect due to $\Lambda$ on traveling gravitational waves in 
	Friedmann universes}
\author{
Stefano Viaggiu,\\
Dipartimento di Matematica,
Universit\`a di Roma ``Tor Vergata'',\\
Via della Ricerca Scientifica, 1, I-00133 Roma, Italy.\\
E-mail: {\tt viaggiu@axp.mat.uniroma2.it}}
\date{\today}\maketitle

\begin{abstract}
In this paper we continue the investigation concerning the propagation of gravitational waves in a cosmological 
background using Laplace transform \cite{A}. We analyze the possible physical consequences 
of the result present in \cite{A} where it is argued that 
a non-vanishing positive abscissa of convergence caused by the de Sitter expansion factor $a(t)=e^{Ht}$ 
implies a shift in the frequencies domain of a traveling gravitational waves as measured by a comoving
observer. In particular, we show that in a generic 
asymptotically de Sitter cosmological universe this redshift effect does also arise. Conversely, in a universe expanding with, for example, a power law expansion, this phenomenon does not happen.
This physically possible new redshift effect, although negligible for the actual very low value of $\Lambda$, can have
interesting physical consequences concerning for example its relation with Bose-Einstein condensation or more speculatively with
the nature of the cosmological constant in terms of gravitons, as recently suggested in \cite{B} near a Bose-Einstein condensation phase.
\end{abstract}
PACS Number(s): 04.30.-w, 04.20.-q, 05.20.-y, 05.30.Jp

\section{Introduction}

The recent \cite{1} detection of gravitational waves from black holes merging certainly represents a milestone in modern physics
opening the possibility to study the features of a given source in terms of its quasi-normal modes. 

As well known \cite{1a,1b}, in order to study the
evolution of fluctuations in the primordial inflation, Fourier transform
is used in a cosmological context but in the wavelengths domain and using a conformal time instead of the
comoving one. However, our interest is not 
focused on the study and evolution of primordial fluctuations but rather on the issue concerning propagation of 
GW generated by a given astrophysical source in a cosmological context but in frequencies domain, i.e. we are mainly 
interested on possible effects on the frequency of a traveling GW caused by the expansion of the universe as measured by
a comoving observer.

In the usual 
mathematical treatment of gravitational waves \cite{2}-\cite{9}(GW thereafter)
emitted by a generic astrophysical source, Einstein's equations are perturbed and thus linearized about a 
given asymptotically flat stationary background. The background time independence (time traslational invariance)
allows one to use Fourier transform to study the
modes of propagation of GW as measured by a distant observer at spatial infinity. This method works very well to study the 
quasi-normal modes \cite{2,5,6,7,8,9} emitted by an isolated source. Unfortunately, we live in an expanding time-dependent
non asymptotically flat universe and the usual technique in terms of Fourier transform in 
frequencies domain is no longer available.
Also the presence of the cosmological scale factor $a(t)$ in the unperturbed metric takes the perturbed one no longer
Fourier integrable. Moreover, issues arise for the formulation of the mathematical framework suitable to obtain
asymptotic conditions depicting
GW in a cosmological context: to this purpose, see the interesting papers in \cite{12,13,14,15} where this issue is addressed 
in a de Sitter universe. In \cite{16,17} the authors obtain and study the axial and polar perturbation equations to test
Huygens principle in Friedmann universes by using expanding modes in wavelengths domain. With the exception of the work
in \cite{18}, where the explicit knowledge of Bondi-Sachs coordinates in a de Sitter universe makes possible a Fourier study in the frequencies domain, no attempts are present in the literature in order to generilize the usual approach in terms of tensorial
spherical harmonics in frequencies domain to a cosmological context. To this purpose, in \cite{A} we have proposed to use
Laplace transform instead of Fourier transform. The Regge-Wheeler equation has been obtained in a de Sitter expanding
universe by means of Laplace transform with respect to the comoving time $t$, the time we measure within our galaxy.
As a consequence, a coupling between the Laplace transformed Regge-wheeler function $Z(s,r)$ and its translation with respect to
the Laplace complex parameter $s$, namely
$Z(s-2H,r)$ (with $H$ the Hubble flow), is present. In \cite{A} it is argued that to a non-vanishing positive abscissa of convergenge
can be associated a possible new cosmological redshift effect in terms of a shift in the frequency emitted by a given source 
as a function of the cosmological constant $\Lambda$. In the present note, we explore some possible physical consequences 
related to this new potential effect.

In section 2 we present our approach in a generic Friedmann universe. In section 3 we derive a new general formula containing the
usual cosmological redshift formula with the new effect. In section 4 we study the relation between the supposed redshift
effect and a Bose-Einstein condensate fluid together with the relation with the cosmological constant. Finally, section 5
is devoted to some conclusions and final remarks.

\section{Laplace transform in a cosmological context}

To start with, we consider a generic Friedmann spacetime expressed in terms of usual comoving coordinates
\begin{equation}
ds^2=c^2dt^2-a^2(t)\left(\frac{dr^2}{1-kr^2}+r^2\sin^2\theta\;d{\phi}^2+r^2d{\theta}^2\right),
\label{1}
\end{equation}
with $k=\{-1,0,1\}$. In \cite{A} we considered the spatially flat de Sitter solution with $k=0,a(t)=e^{Ht}$ and
$H=c\sqrt{\Lambda/3}$ but the technique depicted in \cite{A} can obviously be extended to any cosmological spherically 
symmetric spacetime. As usual, with 
$g_{ik}^{(0)}$ we denote the unperturbed metric (\ref{1}) and with 
$h_{ik}$ a small perturbation $|h_{ik}|<<|g_{ik}^{(0)}|$ where
\begin{equation}
g_{ik}=g_{ik}^{(0)}+h_{ik}.
\label{2}
\end{equation} 
The spherical symmetry of $g_{ik}^{(0)}$ allows one to express $h_{ik}$ in terms of a basis of spherical tensorial harmonics
that in turn are functions of
spherical Legendre polynomials 
$Y_{\ell m}(\theta,\phi), \ell\in\mathbb{N}, m\in\mathbb{Z}, m\in [-\ell, +\ell]$.
As well known, perturbations are classified with respect to the behaviour of the spherical tensorial basis
under parity operator: the polar perturbations look like ${(-1)}^{\ell}$ while the axial ones look like
${(-1)}^{\ell+1}$. In the diagonal gauge \cite{5,6,7,8,9} the polar perturbations depend on 
four functions $N(t,r),L(t,r),T(t,r),V(t,r)$ and the axial ones depend on two functions
$h_0(t,r),h_1(t,r)$. By denoting with
\begin{eqnarray}
X_{\ell m}(\theta,\phi) &=& 2Y_{\ell m,\theta,\phi}-2\cot\theta\;Y_{\ell m,\phi}\label{3}\\
W_{\ell m}(\theta,\phi) &=& Y_{\ell m,\theta,\theta}-\cot\theta\;Y_{\ell m,\theta}-
\frac{1}{\sin^2\theta}\;Y_{\ell m,\phi,\phi}\nonumber\\
H_{11}(t,r,\theta,\phi) &=&TY_{\ell m}+\frac{V}{\sin^2\theta}\;Y_{\ell m,\phi,\phi}+
V\cot\theta\;Y_{\ell m,\theta}\nonumber\\
H_{33}(t,r,\theta,\phi) &=&TY_{\ell m}+VY_{\ell m,\theta,\theta},
\end{eqnarray} 
the line element becomes
\begin{eqnarray}
ds^2 &=& c^2dt^2-a^2(t)\left(r^2\sin^2\theta\;d{\phi}^2+\frac{dr^2}{1-kr^2}+r^2 d{\theta}^2\right)+\label{4}\\
&+&\sum_{\ell m}\Bigl\{2c^2N(t, r) Y_{\ell m}dt^2-
2\frac{a^2(t)}{(1-kr^2)}L(t,r) Y_{\ell m}dr^2-\nonumber\\
&-& 2a^2(t)r^2\sin^2\theta H_{11}(t, r,\theta,\phi)d{\phi}^2-2a^2(t)r^2H_{33}(t, r,\theta,\phi)d{\theta}^2+\nonumber\\
&+& 2ch_0(t,r)\sin\theta\;Y_{\ell m,\theta}\;dt d\phi-
2ch_0(t,r)\frac{Y_{\ell m,\phi}}{\sin\theta}dt d\theta+\nonumber\\ 
&+& 2h_1(t,r)\sin\theta\;Y_{\ell m,\theta}\;dr d\phi-
2h_1(t,r)\frac{Y_{\ell m,\phi}}{\sin\theta}dr d\theta-\nonumber\\
&-& \left[4r^2 V(t,r) Y_{\ell m,\theta,\phi}-4r^2 V(t,r)\cot\theta\;Y_{\ell m,\phi}\right]d\theta d\phi\Bigr\}.\nonumber
\end{eqnarray}
In the standard procedure the perturbed metric in (\ref{4}) is Fourier transformed with respect to the time coordinate.
Unfortunately, the time dependence of the unperturbed metric with the scale factor $a(t)$ makes the line element
(\ref{4}) no longer Fourier integrable. Moreover, if we consider a universe with a big bang singularity, the expansion modes in the frequency domain cannot be extended as in the Fourier case for $t\in(-\infty,+\infty)$. 
To overcome these issues, as argued in \cite{A},
to study the perturbed metric in frequencies domain, an expansion mode using Laplace transform with respect to the 
comoving time $t$ (the time we use within our comoving galaxy) can be performed. For a generic metric perturbation term
$A(t,r)=\{N,L,T,V,h_1,h_0\}$ we have
\begin{eqnarray}
& & a^2(t) A(t,r)=\frac{1}{2\pi\imath}\lim_{b\rightarrow +\infty}\int_{s_0-\imath b}^{s_0+\imath b}A_a(s,r)e^{st}ds,\nonumber\\
& & A(t,r)=\frac{1}{2\pi\imath}\lim_{b\rightarrow +\infty}\int_{s_0-\imath b}^{s_0+\imath b}A(s,r)e^{st}ds,\nonumber\\
& & A_a(s,r)=\int_{0}^{\infty}a^2(t)A(t,r)e^{-st}dt,\nonumber\\
& & A(s,r)=\int_{0}^{\infty}A(t,r)e^{-st}dt.\label{5}
\end{eqnarray}
where $s_0>a_0\geq 0$ where $a_0$
denotes the abscissa of convergence. Fortunately, for our purposes it is not necessary to write down the linearized
Einstein's equations for the expansion modes (\ref{4})-(\ref{5}). Generally, linearized equations for both polar and axial perturbations are\footnote{See the paper \cite{A} for the de Sitter case} a complicated mixture of $A(s,r),A_a(s,r)$
and their derivatives with respect to the radial 
coordinate $r$. However, by inspection of (\ref{4}) and using the (\ref{5}) we can obtain a condition for the existence of the Laplace transformed perturbation $h_{ik}$. By definition $|\{A(t,r)\}|<<1$ but cannot
be generally Laplace transformed with $s=-\imath\omega$, i.e. on the imaginary axis. Moreover, the existence of 
$A_a(s,t)$ generally depends on the asymptotic behavior of $a(t)$. In particular, for $a(t)$ such that 
$a(t)< Me^{\alpha t}$ with $M\in\Re$ and $\forall \alpha\in{\Re}^+$, generally we
have that $a_0=0$ and as a result we have the convergence
of Laplace transform $L(a^2(t)A(t,r))(s)$ for 
$\Re(s)\geq 0$. For when $a(t)\sim e^{Ht}$ (de Sitter expansion \cite{A}) we have $a_0=2H$ and consequently
we have convergence for $\Re(s)>2H$. Finally, if $a(t)> Me^{\alpha t}$ with $M\in\Re$ and
$\forall \alpha\in{\Re}^+$ then Laplace transform $L(a^2(t)A(t,r))(s)$ cannot exist.\\
The abscissa of convergence $a_0$, as explained in \cite{A}, can have a nice physical interpretation. In fact, consider 
for simplicity an axial GW propagating in a Minkowski space ($k=0$ and $a(t)=1$ in (\ref{4})). With
$h_1(s,r)=rZ(s,r)$ the Regge-Wheeler Laplace transformed equation it gives
\begin{equation}
Z_{,r,r}(s,r)-\frac{\ell(\ell+1)}{r^2}Z(s,r)-s^2Z(s,r)=0.
\label{6}
\end{equation}
Since in a Minkowski spacetime $h_{ik}$ is both Fourier and Laplace transformable, and since 
$s=-\imath\omega+s_0, s_0\geq 0$, the perturbations are Fourier transformable on the imaginary axis $s_0=0$ and as a result
the usual expression using Fourier transform is obtained with $s=-\imath\omega$. Hence, in the Minkowski case
we have ${\omega}^2=\lim_{\Re(s)\rightarrow 0}(-s^2)$. Unfortunately, for the presence $a(t)$, the perturbed metric 
(\ref{4}) is no longer Fourier transformable on the complex axis $a_0=0$
and a direct comparison with Fourier transform is impossible.
The problem in a cosmological context is how to interpret the complex parameter $s$. In \cite{A} it is
argued that a positive abscissa of convergence can generate a cut in the frequency of a 
traveling GW as perceived by a comoving observer.
In practice, a mode of frequency $\omega$ is perceived (measured) by the comoving observer with the shifted value
${\omega}_p$ given by
\begin{equation}
{\omega}_p^2=\Re\left(\lim_{\Re(s)\rightarrow a_0}-s^2\right)=\omega^2-a_0^2.
\label{7}
\end{equation}
According to this interpretation for $a_0$, all modes with frequancy $\omega>a_0$ are redshifted. 
Modes with angular frequency $\omega<a_0$ do not propagate. Finally, modes with $\omega\sim a_0$ are perceived as 'frozen'
by the comoving observer. As noticed in \cite{A} these behaviors are in agreement with the usual results concerning tensorial
(i.e. GW modes) perturbations during primordial inflation in wavenumbers space ${\bf k}$ where
modes with $k/a<<H/c$ are practically frozen, while 
modes with $k/a(\eta)>>H/c$ 
are left unchanged by the expansion of the de Sitter universe.\\ 
For a de Sitter universe we have
$a_0=2H$ and by considering modes composed of massless gravitons
with proper wavenumber ${\bf k}/a$  and denoting the proper wavelength as
measured by the comoving observer with ${\lambda}_p$, the (\ref{7}) becomes
\begin{equation}
{\omega}_p^2=\omega^2-\frac{4}{3}\Lambda c^2=\frac{4\pi^2}{{\lambda}_p^2}.
\label{9}
\end{equation} 
From the (\ref{9}) it follows that Hubble horizon modes follow for $\omega\sim H$, while hight under Hubble horizon modes 
are obtained for $\omega>>2H$.\\
Physically, only for cosmological modes evolving, as the de Sitter one or that are asymptotically de Sitter as the actual
concordance cosmological model where
\begin{equation}
a(t)={\left(\frac{{\Omega}_m}{{\Omega}_{\Lambda}}\right)}^{\frac{1}{3}}{\sinh}^{\frac{2}{3}}\left(\frac{3}{2}H_0 t
\sqrt{{\Omega}_{\Lambda}}\right),
\label{10}
\end{equation}
with $k=0$, ${\Omega}_m\simeq 0.32$, ${\Omega}_{\Lambda}\simeq 0.68$ and $H_0$ the present day Hubble flow, formula
(\ref{9}) is valid with $a_0=2c\sqrt{\Lambda/3}$. For example, for power law cosmologies with $k=0$ and
$a(t)\sim t^{\alpha},\;\alpha\in{\Re}^+$, the frequency spectrum as emitted by a given astrophysical source is left unchanged
by the expansion of the universe. Conversely, in a universe expanding with $a(t)>Me^{\alpha t},M\in{\Re}^+$ and 
$\forall\alpha\in{\Re}^+$ we have that no mode can propagate as a wave and a GW rapidely vanishes. These results are in
agreement with physical intuitions: it is resonable to think that in a spacetime very quickly expanding, more than a se Sitter 
one, no GW can propagate as a typical wave.\\  
Although for the very low value of $\Lambda\sim 10^{-18}Hz$
the shift effect is expected very small, interesting consequences can arise when the (\ref{9}) does apply.
Some interesting consequences are discussed in the following sections.

\section{A new cosmological redshift formula}

As well known, the wavelength of photons propagating in an expanding universe is stretched by an amount depending from the scale factor $a(t)$, i.e. the cosmological redshift. This phenomenon can also be applied to (massless) gravitons traveling in an 
expanding universe. The cosmological redshift is depicted in terms of the dimensionless quantity $z$: by denoting with
${\omega}_e$ the angular frequency of the emitted radiation and with ${\omega}_o$ the one observed at present time, we have
$1+z=\frac{{\omega}_e}{{\omega}_o}$. In a Friedmann cosmological context we have
\begin{equation}
1+z=\frac{a(t_o)}{a(t_e)},
\label{11}
\end{equation}
where $t_e$ is the emission time of the radiation and $t_o$ the detection time. Formula (\ref{11}) is valid for any Friedmann
universe and the frequencies are measured with respect to a comoving observer. The reasonings of this section are valid only for 
a de Sitter universe with $a(t)=e^{Ht},k=0$ or for the asymptotically de Sitter metric (\ref{10}) or for a universe equipped with
$\Lambda$ and non-zero spatial curvature.\\
To apply formula (\ref{11}), we assume a graviton
with comoving wavelength ${\overline{\lambda}}_c$ and emitted pulsation ${\omega}_e$. However, formula (\ref{9}) implies that
the expanding universe shifts the frequency emitted by a given source by an amount depending on $\Lambda$. Hence, the usual 
redshif formula must be applied to the pulsation ${\omega}_p$, measured with respect to the comoving time $t$, that in turn measures
a comoving wawelengths ${\lambda}_c$ and proper one ${\lambda}_p$ with ${\lambda}_p=a(t){\lambda}_c$:
\begin{equation}
{\omega}_p=\sqrt{{\omega}^2-\frac{4}{3}\Lambda c^2}=\frac{2\pi c}{{\lambda}_p}.
\label{12}
\end{equation}
For a traveling GW it is natural to assume that the comoving wavelength ${\lambda}_c$ is constant in time since no dissipative effects arise.
We thus have for ${\omega}_p$
\begin{equation}
\frac{{\omega}_p(t_o)}{{\omega}_p(t_e)}=\frac{a(t_e)}{a(t_o)}=\frac{1}{1+z}.
\label{13}
\end{equation}
To obtain a compact formula taking into account the usual cosmological redshift and the new one provided by (\ref{12}), we can
put the (\ref{13}) in (\ref{12}) to obtain
\begin{equation}
{\omega}^2(t_o)=\frac{{\omega}^2(t_e)}{{(1+z)}^2}+\frac{4}{3}\Lambda c^2
\left[1-\frac{1}{{(1+z)}^2}\right]. 
\label{14}
\end{equation}
The pulsation ${\omega}(t_e)$ is the one emitted by a given astrophysical source but not the one measured by 
a comoving observer. The formula (\ref{14}) shows that the usual cosmological redshift relation between ${\omega}(t_e)$
and ${\omega}(t_o)$ breaks down: the usual relation is obtained in (\ref{13}) with ${\omega}_p$.  
Following our interpretation for $a_0$, the cosmological constant acts as a shift term in frequencies domain.
Since of the actual very small value for $\Lambda$ this effect is expected to be very small. Nevertheless, 
some interesting physical implications follow.

To our purposes , we introduce a new parameter, namely $\overline{z}$, that is the redshift parameter calculated 
with respect to ${\omega}(t_e)$ and ${\omega}(t_o)$:
\begin{equation}
\frac{{\omega}(t_e)}{{\omega}(t_o)}=1+\overline{z},
\label{15}
\end{equation}
that inserted in (\ref{14}) it gives
\begin{equation}
\frac{1}{{(1+\overline{z})}^2}-\frac{1}{(1+{z}^2)}=
	\frac{4\Lambda c^2}{3{\omega}^2(t_e)}\left[1-\frac{1}{{(1+z)}^2}\right].
\label{16}
\end{equation}
Expression (\ref{16}) shows that the difference is proportional to $\Lambda$ and as a consequence for usual astrophysical sources
with ${\omega}(t_e)\sim 10^{-2}-10^{3}\;Hz$ we have practically $z\simeq\overline{z}$. However, for gravitational waves with a very 
small pulsation ${\omega}(t_e)<<10^{-2}Hz$ the effect is no longer negligible. Hence, these reasonings can be of interest for the 
eventual detection of a stochastic background of GW.

\section{Gravitons and a Bose-Einstein condensate}

To start with, consider a graviton generated long time ago with $z\rightarrow\infty$. From (\ref{14}), we see that, independently
from the value of the emitted angular frequency ${\omega}(t_e)$ we have
\begin{equation}
\lim_{z\rightarrow +\infty}{\omega}(t_o)=\frac{2}{\sqrt{3}}c\sqrt{\Lambda}=2H.
\label{17}
\end{equation}
From the (\ref{12}) it follows that, independently from the value of the emitted frequency ${\omega}(t_e)$ or the one 
effectively measured ${\omega}_p(t_e)$, the observed asymptotic pulsation ${\omega}_p$ is an universal value,
i.e. ${\omega}_p\rightarrow 0$. A graviton emitted in an early primordial cosmological era,
with a given frequency will be perceived asymptotically frozen by a comoving experimenter living in a 
(asymptotically) de Sitter spacetime. Also note that the existence condition for the (\ref{12}) is
$\omega\geq 2c\sqrt{\Lambda/3}$. This implies that the cosmological constant cuts pulsations such that
$\omega < 2c\sqrt{\Lambda/3}$ that cannot propagate for the comoving experimenter. 
Moreover, angular frequencies emitted exactly
with $\omega=2c\sqrt{\Lambda/3}$ are frozen with respect to the comoving observer. In a Friedmann flat universe with 
apparent horizon located at the proper areal radius $L_h=c/H$, the so frozen modes have wavelength ${\lambda}_p$ with
${\lambda}_p\sim L_h$. Hence, modes with $\omega=2c\sqrt{\Lambda/3}$ are in a Bose-Einstein condensation state
(BEC) as seen by a comoving observer. As an example,
we may think to a stochastic background \cite{19} of gravitational waves emitted in an early era
during or soon after primordial inflation or for $t\rightarrow-\infty$ in a de Sitter universe with frequancy 
$\sim 10^{-9}Hz$ or less. Formula (\ref{14}) implies that at present time ${\omega}_o\simeq 2H$ and consequently
${\omega}_p(t_o)\simeq 0$, i.e. gravitons are perceived by a comoving observer as practically frozen in a BEC state.\\
More speculatively, since it is the gravitational constant $\Lambda$ that could generate this
frequency shift (a cut) with respect to the frequency emitted
by a given astrophysical source, it is natural to wonder if this possible effect can hint something about the nature of the cosmological constant and in particular its very small value. First of all, the fact that only in a de Sitter or asymptotically de
Sitter spacetime the supposed frequency-shift effect does arise, according to our interpretation of the abscissa of convergence, could be a hint that only the cosmological constant can 
interact with a traveling GW, i.e. the cosmological constant can be made of massless excitation modes interacting with gravitons 
composing a GW. In effect, an electromagnetic wave propagating in a medium, since of the interactions between photons and electrons
composing the atoms of the medium itself, can slow down its effective phase velocity\footnote{In a medium with index of refraction
	$\eta$, we have $v_f=c/\eta$.} ${v}_f$. In \cite{B}, 
a statistical description of the cosmological constant in terms of massless bosons
(gravitons) has been proposed. By means of usual techniques of statistical mechanics, in \cite{B} it has been shown that in a de Sitter universe, under the assumption that gravitons eventually composing $\Lambda$
have a mean $T_h$ temperature given by the one \cite{VV} of the apparen horizon 
$T_h=\frac{c\hbar}{4\pi k_B L_h}$ and that the phase velocity $v_f$ is slowed down by an amount proportional to 
$1/N^{1/3}$, with
$N$ the number of gravitons within the apparent horizon $L_h$, an effective statistical representation for $\Lambda$ can be proposed.
Moreover, thanks to the very low value of $\Lambda$, these gravitons in \cite{B} has been supposed to be near a BEC state.\\
The result of this paper can provide a simple physical mechanism to create gravitons near a BEC state, i.e. the expansion of the universe.\\
Independently on the viability of the hypothesis presented in \cite{B}, consider an early primordial era where a stochastic
background of GW is created in a spatially flat universe with measured pulsation ${\omega}_p(t_e)$ by the comoving experimenter of the order \cite{19} of
${\omega}_p(t_e)\sim 10^{-9}Hz$. With the help of (\ref{15}) we have 
${\omega}_p(t_o)={\omega}_p(t_e)/(1+z)$. By supposing for example that the universe before nucleosynthesis ($t\sim 1 sec.$) has been 
composed with radiation ($a(t)\sim t^{1/2}$), then a value for ${\omega}_p(t_o)\sim 10^{-18}Hz$, i.e. such that
${\omega}_p\sim c\sqrt{\Lambda}$ near a BEC state, is obtained for $z\sim 10^9$ corresponding to $t_e\sim 10^{-18}sec.$
Conversely, by supposing a universe filled with stiff matter ($a(t)\sim t^{1/3}$) before nucleosynthesis, we found
$t_e\sim 10^{-27}sec$, soon after the primordial inflation. Hence, a 
stochastic background of GW emitted after primordial inflation and before nucleosynthesis can be perceived as practically frozen by a present day comoving observer forming a condensate near a BEC state.\\
Notice that the phenomenon depicted above does apply to the perceived pulsation ${\omega}_p$ rather than the usual one
$\omega$ and this represents a fundamental physical difference with the usual 
redshift effect. To this purpose, consider a given astrophysical source in a de Sitter universe emitting GW at frequencies ${\omega}(t_e)$ ranging
from $\omega(t_e)\in({\omega}_1,{\omega}_2)$ with ${\omega}_1<2H$ and ${\omega}_2>2H$. Formula (\ref{12}) implies that only frequencies emitted with $\omega >2H$ can propagate (redshifted). As a consequence, frequencies with $\omega(t_e)\simeq 2H$ are 
practically in a BEC state and the ones with $\omega(t_e)=2H$ are frozen in a BEC state and as a result we have
$\omega(t_e)=2H,\;\forall t\geq t_e$, i.e. ${\omega}_p=0,\;\forall t\geq t_e$. If the reasonings above take effectively place 
in the real universe, we have depicted a possible cosmological mechanism to obtain massless gravitons in a BEC state.
In ordinary physics, generally BEC cannot arise for massless bosons. However, quite recently the possibility that photons can condensate has been experimentally \cite{37} obtained. In practice, 
thanks to the 'finite' dye-filled optical microcavity where experiment is performed,
photons acquire an effective mass near the BEC ground state. If the reasonings above are correct, the phenomenon presented in this paper could be a 'cosmological' realization for gravitons BEC where the 'dye' is the cosmological constant and the cavity is
provided by the Hubble horizon \cite{B,VV} at $L_h=c/H$.

Summarizing, the shift effect depicted in this paper is expected to be very small thanks to the very small value for 
$\Lambda$. Nevertheless, interesting physical consequences can arise for GW traveling in a de Sitter or asymptotically de Sitter
universe. In particular, very low frequencies emitted with ${\omega}=2H$ are 
frozen in a BEC-like state with respect to a comoving observer. This can be certainly matter for further investigations.

\section{Conclusions and final remarks}

In this paper we have further investigated the hypothesis present in \cite{A} that the non vanishing abscissa of convergence can generate on a monochromatic GW with an emitted pulsation $\omega(t_e)$. From a mathematical point of view, 
in a cosmological time dependent spacetime we are forced to use Laplace transform instead of Fourier one since of the lack of
time-translational invariance of the background and on the presence of $a(t)$, making the metric coefficients generally no more
integrable. However, the shift phenomenon given by the (\ref{12}), following our reasonings, does appear only in a de Sitter
universe equipped with a positive $\Lambda$ or asymptotically de Sitter (perhaps the one we live). In a spacetime with, for example,
$a(t)\sim t^{\alpha},\alpha\in{\Re}^+$, we have $a_0=0$ and, thanks to the formula (\ref{7}), formula (\ref{12}) does not apply.
In practice, the cosmological constant is acting as a kind of damping parameter $\zeta$ 
in ordinary dynamical systems. In fact, by denoting with $\omega$ the natural angular frequency (i.e. the angular
frequency without $H$) of a monochromatic GW and with $2H=\zeta\omega$, 
the one effectively measured ${\omega}_p$ is given by ${\omega}_p=\sqrt{{\omega}^2-4H^2}=\omega\sqrt{1-{\zeta}^2}$.
In this regard, the cosmological constant acts as an elastic material damping the frequency of the traveling GW.\\
As a consequence, a GW with a 'natural' emitted frequency ${\omega}=2H$, will be seen as frozen by a comoving observer in
a de Sitter universe. In practice, gravitons emitted with natural frequency given by $\omega=2H$ are damped by 
$\Lambda$ up to a BEC state. If this phenomenon would be effectively realized in the real universe, it will be 
interesting to compare this situation with the one concerning BEC for photons. In \cite{37}, it has been realized that photons can effectively condensate
within a dye-filled optical microcavity. Photons so obtained acquires an 'effective' mass within the cavity and this takes possible the condensation phenomenon. 
In our background, consider equation (\ref{12}) multiplied by the Planck constant $\hbar$. We have:
\begin{equation}
{\hbar}^2{\omega}^2={\hbar}^2{\omega}_p^2+\frac{4}{3}{\hbar}^2c^2{\Lambda}.
\label{18}
\end{equation}
Following the reasonings above related to the BEC, we can see the (\ref{18}) as an effective dispersion relation satisfied by the gravitons $E^2=c^2p^2+m_{eff}^2 c^4$. Hence, $\hbar{\omega}_p=cp$ and 
$m_{eff}=\frac{2\hbar\sqrt{\Lambda}}{\sqrt{3}c}$. Numerically we have $m_{eff}\simeq 10^{-65}g$. We can also obtain a very crude estimation for the critical temperature of a BEC made of gravitons. In fact, suppose to have a stochastic background of GW in
a de Sitter universe and suppose that gravitons within the apparent horizon $L_h=c/H$ are thermalized. Inside  the apparent horizon $V_h=4/3\pi L_h^3$ we have, at zero chemical potential $\mu=0$ for the bose gas with 'effective' mass $m_{eff}$ and using a non-relativistic approximation for $E^2$ \cite{LL}:
\begin{equation}
T_c=\frac{3.31}{2^{\frac{2}{3}}}\frac{{\hbar}^2}{m_{eff}}{\rho}^{\frac{2}{3}},\;\;\rho=\frac{N}{V_h},
\label{19}
\end{equation}
where $N$ are the gravitons within $L_h$. As a consequence of (\ref{19}) we obtain 
$T_c\sim N^{\frac{2}{3}}c\hbar\sqrt{\Lambda}$. Hence $T_c\sim N^{\frac{2}{3}}10^{-29}K$. As a consequence, for a huge value for 
$N$, for example $N=10^{90}$, we have a very huge value for $T_c$. However, since we expect a stochastic background of GW with a low frequency ($\leq 10^{-9}Hz$, see \cite{19}), a reasonable value for $N$ could be $N<10^{30}$ with $T_c>10^{-29}K$. In any case,
since with the cosmological constant $\Lambda$ can be associated \cite{VV} the temperature of its apparent horizon
\footnote{Coinciding with the Hubble radius in a de Sitter universe.} $T_h\sim c\hbar\sqrt{\Lambda}$, for gravitons thermalized
with $\Lambda$, we always have $T_h<T_c$. This crude estimation shows that gravitons of a stochastic 
GW background are at present time in (or near) a BEC state can be a viable possibility.

\end{document}